# Topological feature study of slope failure process via persistent homology-based machine learning

Shengdong ZHANG[1] Shihui YOU*[1] Longfei CHEN[2] Xiaofei LIU[2]

1. College of Mechanical and Electrical Engineering, Zaozhuang University 277160;

2. College of Civil Engineering and Mechanics, Xiangtan University 411105

**Abstract:** Using software UDEC to simulate the instability failure process of slope under seismic load, studing the dynamic response of slope failure, obtaining the deformation characteristics and displacement cloud map of slope, then analyzing the instability state of slope by using the theory of persistent homology, generates bar code map and extracts the topological characteristics of slope from bar code map. The topological characteristics corresponding to the critical state of slope instability are found, and the relationship between topological characteristics and instability evolution is established. Finally, it provides a topological research tool for slope failure prediction. The results show that the change of the longest Betti 1 bar code reflects the evolution process of the slope and the law of instability failure. Using discrete element method and persistent homology theory to study the failure characteristics of slope under external load can better understand the failure mechanism of slope, provide theoretical basis for engineering protection, and also provide a new mathematical method for slope safety design and disaster prediction research.

**Key words:** Slope; Persistent homology; Bar code map; Feature selection; Topological analysis

## 1 Introduction

Slope failure exists widely in many fields such as geotechnical engineering, environmental science, and transportation. Slope will be unstable due to rain erosion, earthquake and impact, which will cause landslides and huge economic losses to the society and huge life threats to the people. Therefore, the research on slope failure has very important theoretical and scientific significance. Experimental and numerical analysis methods are the main research methods for slope instability. Numerical analysis methods include Discrete Element Method (DEM) [1], Finite Element Method (FEM) and so on. The advantage of numerical analysis is that it has low cost, good repeatability, and easy implementation. At present, scholars have done much research on the stability of rock mass and slope using DEM. Fakhimi, Potyondy, Backstron, etc. have studied the deformation and failure characteristics, macroscopic properties and microscopic characteristics of rock masses by the method of numerical simulation and uniaxial compression tests [2-5]. Based on PFC2D and EDEM software, Yang Bing, Zhang Xiaoxue, and Yang Ling have studied and discussed the macro-response of the slope instability process [6-10]. Su Yonghua, Li Shuai, etc. used the double strength reduction method to study the ultimate bearing capacity of the slope [11].Pan Min, Ling Chen et al. proposed a slope reliability analysis method based on the sparse grid collocation method, which provides an effective way for complex slope reliability analysis [12]. The early research on slope failure is carried out by means of experiment, but this method only involves one scale. Slope failure is a dynamic and multi-scale failure process, which involves the accumulation of damage from micro pores to the formation of macro cracks. The micro mechanical evolution is the root of macro mechanical evolution. Macro and micro systems have cross scale correlation, hierarchy and randomness, so we need to find a research method that can cover all scales, and continuous coherence is a topological homology method that can cover continuous scales. Therefore, the persistent homology method can be used to analyze the whole process of slope failure from crack generation to instability failure, and find the topological characteristics corresponding to the critical state of instability, thus providing a new

---

**Corresponding author

**E-mail address:** 101434@uzz.edu.cn

tool for slope failure prediction. At present, Discrete Element Method and other discontinuous deformation methods have been popularized [13], which is not mature in quantitative evaluation of slope stability, and there is no application of topological feature analysis in slope failure process characterization. Therefore, this thesis introduces the continuous coherence theory to describe the topological characteristics of the slope. This is of great engineering and scientific significance. In recent years, persistent homology has made great progress in image processing and target recognition. Zhang Jingliang, Ju xianmeng [14] used the calculation method of persistent homology and simple complex homology group to classify and recognize images. By calculating the homology of the simple complex, the barcode is obtained, and the topological features and corresponding geometric structure information of the image are obtained based on the barcode. The stability of the slope is studied by the discrete element software UDEC, and the failure process of the slope is analyzed, and the mechanical behavior of the slope in different time periods is analyzed. Finally, the topological analysis of the slope is carried out, and the overall safety factor of the slope is obtained by using the strength reduction method of the software to judge the stability. The slope is regarded as a multi-scale structure network composed of various rock blocks, Based on the theory of persistent homology, the topological characteristics of slope are analyzed, and the barcode diagram is obtained. Then, combining the persistent homology theory and support vector machine, the features of slope bar code map are extracted, and the main features related to its category are retained. The relationship between slope instability and topological characteristics is established, and the damage evolution law of slope is described based on the parameters of barcode diagram. The specific implementation process is as follows.

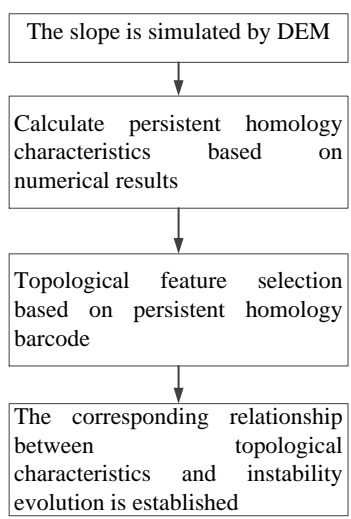

Fig.1 The specific implementation process

Although different slopes may be different in material, state and boundary constraint conditions, they will go through a process from crack generation, accumulation and evolution to final slope failure. If the whole slope is regarded as a network composed of rock blocks, then the evolution law of the network is consistent, that is to say, its inherent topological characteristics are the same. Compared with the traditional instability criteria such as strength reduction method, the results of this thesis are also consistent. Therefore, the research method and results have wide applicability.

## 2 Basic principles of persistent homology
### 2.1 Simplicial Complexes
Simplicial complex is formed by connecting different simplexes, and its boundary is its surface. It is precisely defined as:

Definition: In $R^N$ space, the simple complex satisfies the following definition:

1. Any face of a simplex from $K$ is also in $K$.

2. The intersection of any two simples in *K* is either empty or shares faces.

According to the definition, the intersection of two different simplexes in a simple complex must be a common point. The simple complex in the example below is a two-dimensional simple complex, because its highest dimensional simplex is a two-dimensional triangle.

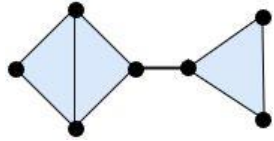

**Fig.2 2-dimensional simple complex**

At the same time, special simplex and simple complex should be distinguished. If it is also a triangle, and one of the three edges is connected, then the simple complex is composed of one-dimensional simplex. And if you fill the inside of the triangle, that is to say, every point in the triangle can be taken, that is, a two-dimensional simplex, which is a simplex. So this highlights the importance of coloring.

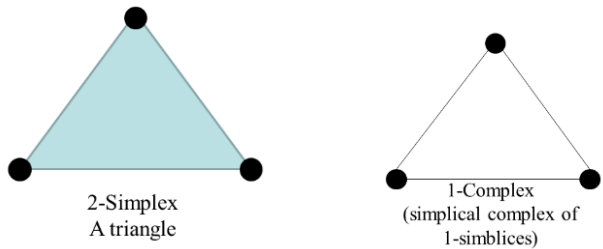

**Fig.3 Color distinction**

The surface of a simplex is its boundary. When describing a simplex or a complex, the traditional "coloring" on the surface of a simplex makes it clear that it is a "solid". As shown in the figure below.

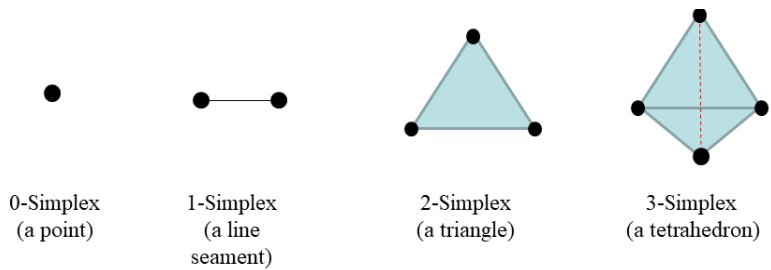

**Fig.4 Each(0-3) dimensional simplex**

## 2.2 Homology group

Group is a kind of mathematical structure, which is used to describe all things with symmetry concept. And now it has developed into a complete mathematical system group theory. Group is a very common concept, it can be applied to many objects. For example, the equilateral triangle in mathematics, which has the axis of symmetry and can return to the original structure after rotation, can be described by groups. Group theory provides a good tool for describing such operations.

The exact definition of a group is given below.

Group: a set *G* and a binary operation defined on it constitute a group. The binary operation maps $a,b \in G$ to $c \in G$, and writes $a*b=c$, $\forall\ a,b,c \in G$. When the set and the defined binary operation satisfy the following three conditions, they are effective groups.

1. Associative law

$$\forall a,b,c \in G,\ (a*b)*c = a*(b*c)$$

2. Identity element: There is an element $e \in G$, after binary operation, the elements $a$ and $e$ in $G$ are still mapped back to the element itself in the set $G$, that is $a*e=e*a=a$. And the element is unique, so we call $e$ is the unit element.

3. Inverse element: If any $a \in G$ has an element $b \in G$, $a*b=b*a=e$, where $e$ is a unit element, then $b$ is the inverse element of $a$

For example, the set of integers and addition operations can form a group. Because of the commutative law of the addition of two numbers $(a+b)+c=a+(b+c)$ and the existence of the identity element 0 in the set of integers, any number added with 0 will get itself. The opposite number of any element is its inverse element, because the sum of the sum of the opposite number and the original number is the unit element 0

Homology group: The k-th homology group is a quotient group $H_k=Z_k/B_k$. The sequence of the kth homology group is called the kth Betti number and satisfies the $\beta_k$=rank $H_k$=rank $Z_k$-rank$B_k$.

Intuitively, $b_0$ is the number of connected parts, $b_1$ is the number of one-dimensional holes, $b_2$ is the three-dimensional holes surrounded by faces, and so on. For example, the Betti number of a torus should be $b_0 = 1$, $b_1= 2$, $b_2= 1$. It represents a connecting component, two one-dimensional holes and one three-dimensional hole.

## 2.3 Persistent homology

Persistent homology is the introduction of persistence on the basis of homology, which is the generalization of homology. Homology is a static concept, and persistent homology is a changing, dynamic process. Homology calculation is carried out on simple complex, while persistent homology is calculated on filtered nested complex, which is a continuous change process. The calculation of persistent homology is closely related to homology group.

## 3 Numerical simulations

### 3.1 Calculation model

The discrete element model is established according to the actual situation of the rock slope behind the powerhouse of Jiangpinghe hydropower station. Due to the large angle of the free face, the weathering unloading cracks are developed. The lowest elevation is about 380 m, and the slope height is 100-120 m, which is vertical to the river. The slope is mainly composed of thick limestone and marl formed by weathering and rain erosion. The horizontal depth of the strongly weathered rock mass is 5-10 m, and the fault and bedding plane incline to the slope at a small angle, so it is unlikely to form an unstable wedge. Among all kinds of structural planes, the unloading fissures with steep dip angle along the slope are developed, and the depth is 1-10 m, which is the controlling structural plane affecting the stability of the slope. In this thesis, referring to the actual slope mentioned above, the discrete element software UDEC is used to simulate the process of slope instability. The discrete element model is shown in Figure 5.

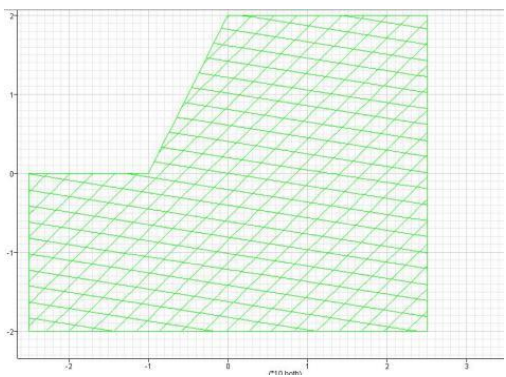

**Fig.5 Rock slope model**

The whole rock slope is set up with 20 meters high, 25 meters wide and 20 meters high. The dangerous rock is equipped with joints in both vertical and horizontal directions, with a spacing of 2 meters. Rock masses can be translated and rotated, while joints can be compressed and separated. Therefore, the model also allows large deformation such as rock block falling off.

The mechanical parameters and Joint mechanical parameters are listed in Table.1 and Table.2.

**Table. 1 Mechanical parameters of rock mass**

| Destiny(kg m$^{-3}$) | 2600 | Internal friction angle(°) | 35 |
|---|---|---|---|
| Bulk modulus(GPa) | 21 | Cohesion(MPa) | 5 |
| Shear modulus(GPa) | 11 | | |

**Table.2 Joint mechanical parameters**

| Normal stiffness (GPa m$^{-1}$) | 20 | Cohesion(MPa) | 5 |
|---|---|---|---|
| Shear stiffness (GPa m$^{-1}$) | 10 | Strength(MPa) | 2 |
| Internal friction angle(°) | 25 | | |

### 3.2 Dynamic response analysis of rock mass

The right and bottom of the dangerous rock are set as viscous contact, and the top and left side are free boundaries without any constraint. Then a cosine seismic wave with the size of 10MPa and frequency of 30Hz is applied at the bottom. Then it is calculated by UDEC software. The following is the change response diagram of rock mass under multiple loading.

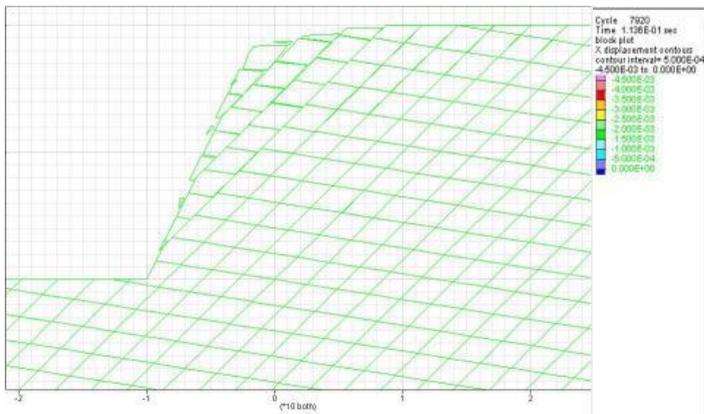 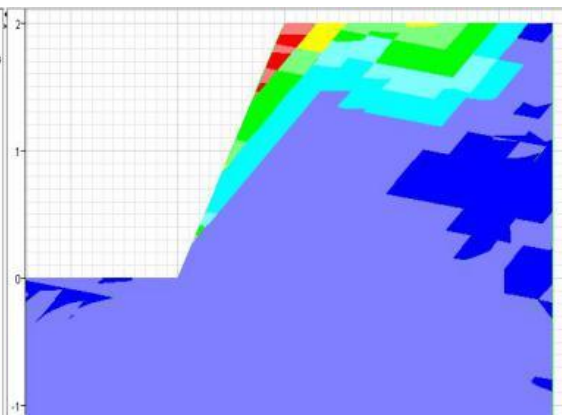

**Fig.6 First load damage diagram**  **Fig.7 First load displacement diagram**

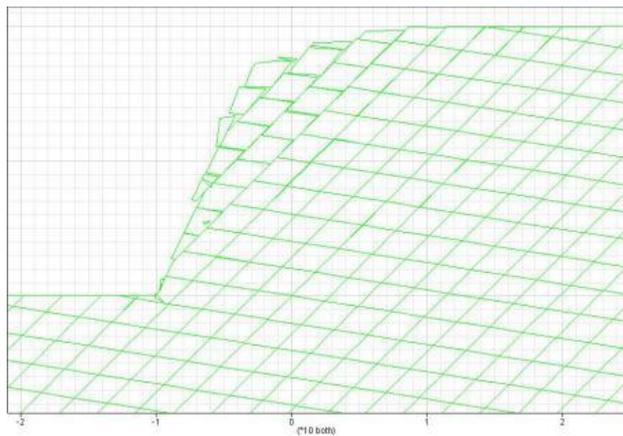 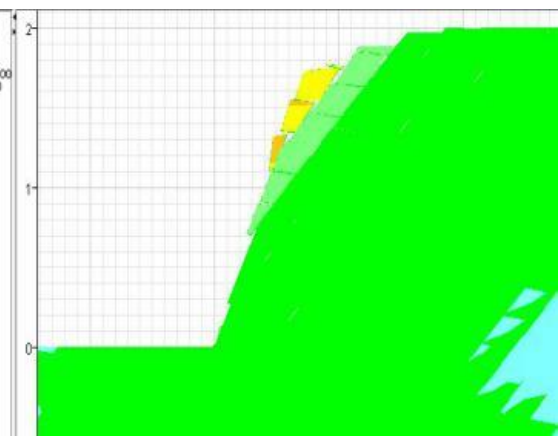

**Fig.8 Second load damage diagram**  **Fig.9 Second load displacement diagram**

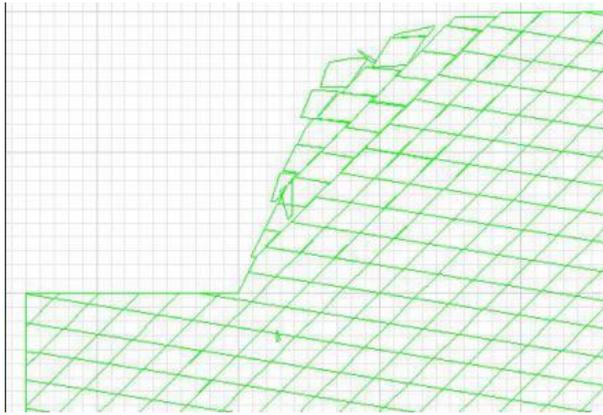 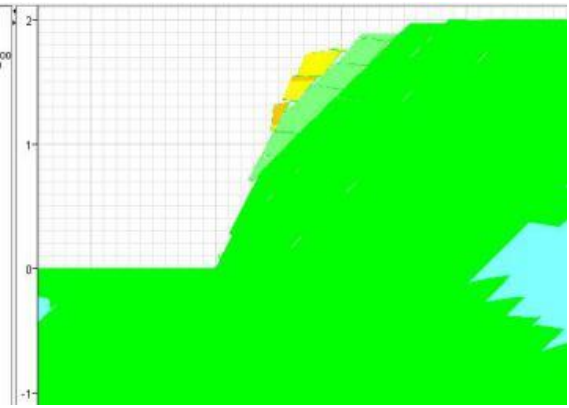

**Fig 10 Third load damage diagram**          **Fig 11Third load displacement diagram**

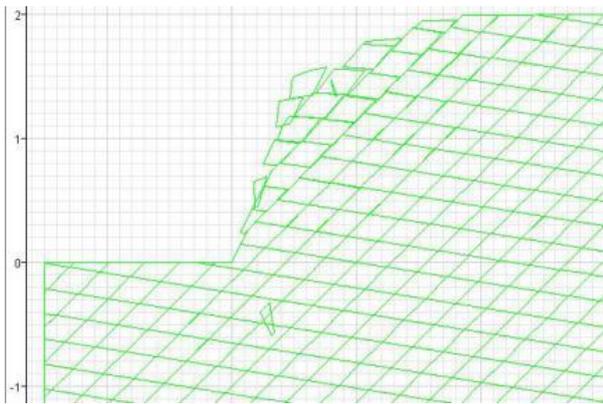 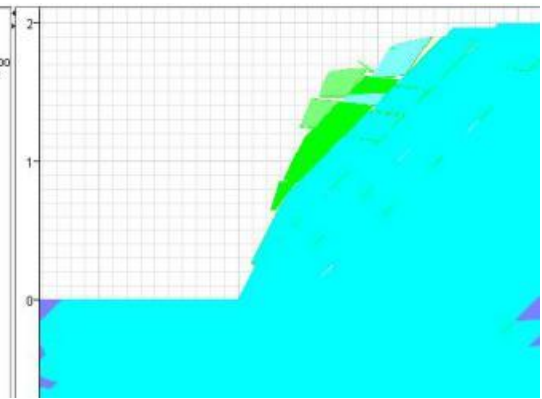

**Fig.12 Fourth load damage diagram**          **Fig.13 Fourth load displacement diagram**

Through the analysis of the failure process, we can see that when the first loading, the three outermost blocks have a slight slip, and the internal rock mass is basically in a stable state due to the constraint of the outer rock mass, and it is relatively close, and there is no large separation. When the second loading, we can see that the outermost rock mass has a large slip, and the top of the inner layer also slightly subsides. Under the third loading, the failure trend of the two layers is further obvious. When the fourth loading is carried out, the slope is basically subject to sliding failure, and the outermost rock blocks fall off one after another. It can be seen from the overall displacement nephogram that the overall displacement of the slope increases with the increase of the loading times, but the displacement growth of the outermost layer is obviously faster than that of the inner side of the slope. It can also be seen that the slip zone of rock mass will be formed.

## 4 Selection of topological features and establishment of correlation between persistent homology and slope failure

### 4.1 feature selection

In order to implement the topological method in support vector machine, 11 bar code graph statistics in reference [15] are used to represent the topological structure information of slope. Different statistics are selected from Betti 0 and Betti 1 barcode information to represent the characteristic information of slope type. The length and position of bar code in literature are in angstroms are shown.in Table.3.

### 4.2 Establishment of correlation between continuous coherence and slope failure

After getting the dynamic failure process of slope under seismic load, the next step is to find the representative moments above, find the centroid coordinates of soil particles, calculate with javaplex, and draw the bar code diagram.

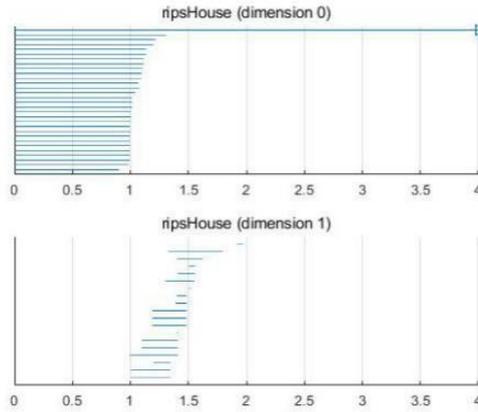 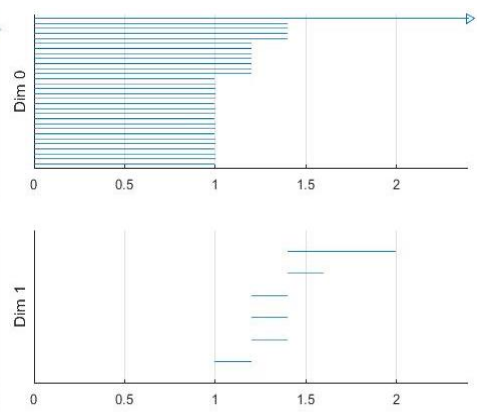

| | Fig.14 First load bar code | Fig.15 Second load bar code |

**Table.3 A list of features used in support vectors**

| Feature | Betti | Description |
| --- | --- | --- |
| 1 | 0 | The length of the second longest Betti 0 bar. |
| 2 | 0 | The length of the third longest Betti 0 bar. |
| 3 | 0 | The summation of lengths of all Betti 0 bars except for those exceed the max fltration value. |
| 4 | 0 | The average length of Betti 0 bars except for those exceed the max fltration value. |
| 5 | 1 | The onset value of the longest Betti 1 bar. |
| 6 | 1 | The length of the longest Betti 1 bar. |
| 7 | 1 | The smallest onset value of the Betti 1 bar that is longer than 0.5. |
| 8 | 1 | The average of the middle point values of all the Betti 1 bars that are longer than 0.5. |
| 9 | 1 | The summation of lengths of all the Betti 1 bars except for those exceed the max fltration value. |
| 10 | 1 | The average length of Betti 1 bars except for those exceed the max fltration value. |
| 11 | 1 | The average of the number of barcodes at 0.25, 0.5, 0.75, 1.0, 1.25, 1.5 |

In the failure diagram of the first loading, it can be seen that the outer rock mass has a small slip, and the internal rock mass is basically in a stable state. By observing the corresponding bar code, we find that most of the 0-dimensional Betty numbers are near 1, and a few are beyond 1. The results show that under the action of seismic load, the outermost rock block is reflexively loose and the displacement increases. If we look at the 1-D Betty number, there are many lines and the duration is short. When the first loading, the rock blocks are relatively close and the surrounding holes are small.

From the bar code diagram of the second loading, it can be seen that the longest Betty 1 is longer than the first loading, and the lines are more sparse. This shows that part of the fine cracks have penetrated and merged, forming a few but wider macro cracks.

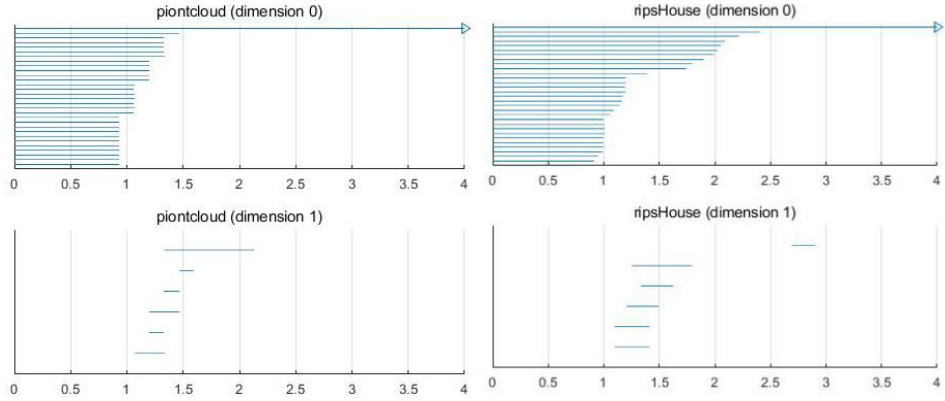

**Fig.16 Third load bar code**     **Fig.17 Fourth loading bar code**

The increase of 0-dimensional Betty number indicates that the rock mass of the slope has been loosened and the displacement has increased. The 1-D Betty number is still sparse. The longest Betty 1 starts at 1.35 and ends at 2.1, and the length is 0.75. The longest Betty 1 length increases. Comparing the two pictures before and after, it can be seen that the end value of Betty 1, which plays a leading role in slope cracks, moves back rapidly and the length also becomes longer. It shows that the width of the main crack which determines the final stability of the slope increases rapidly, which indicates that the slope is in a critical state of instability.

It can be seen from Figure 16 that the bottom of 0-dimensional Betty number barcode increases significantly, and the top basically maintains between 1.5 and 2.5. Compared with the previous figure, the number of 1-D Betty number is basically unchanged, but several lines at the bottom are longer, which indicates that with the fourth load, the number of holes does not change, but the original smaller holes become larger. At the same time, a line with short duration appears in the distance, and the 1D barcode map changes suddenly, which indicates that larger holes are formed in the slope at this time. It shows that the slope has sliding failure at this time.

In this thesis, the four processes of Betty 0 and Betty 1 bar code, select several representative features: feature 5, feature 6, and feature 8. In principle, the feature selection is based on the different emphasis reflected by each feature, combined with the actual engineering characteristics of slope, to select the characteristics that can best reflect its change process and essence. The reason why these three features are selected is that they have a common feature by comparing the topological features mentioned above. They are all related to the cracks of the slope, and the cracks are the most critical factor to determine the stability of the slope. Feature 5 reflects the duration and variation characteristics of the main cracks, feature 6 reflects the size and change trend of the largest main fracture of the slope, and feature 8 reflects the overall crack development level of the slope.

The first is feature 5, whose content is the starting value of the longest Betty one, which reflects the filtering value when the longest ring is formed, that is, the duration from the beginning of the maximum crack formation to the complete failure. The starting values of the longest Betty I in these four moments were 1.35, 1.4, 1.35, 1.2, and the ending values were 1.75, 2, 2.1 and 1.75 respectively (all in the order of increasing time). The line diagram is shown in Fig.18.

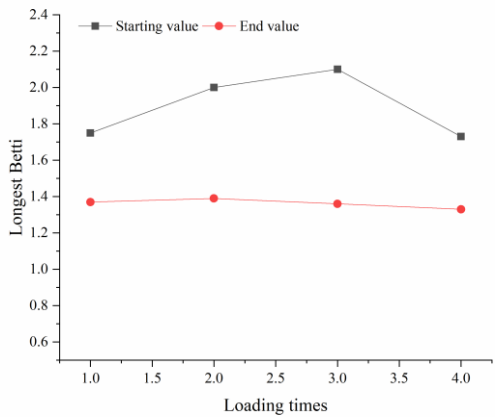 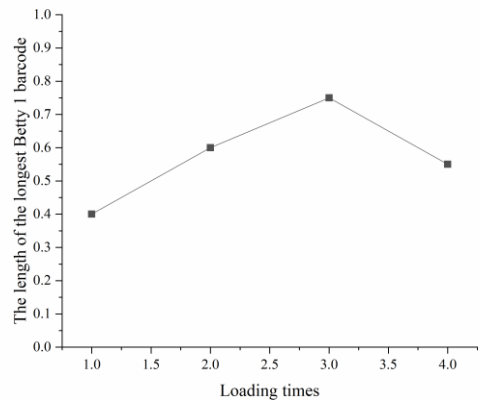

       Fig.18 Line program of feature 5       Fig.19 Line program of feature 6

It can be seen from Fig.18 ,the broken line diagram that both the starting and ending values of the longest Betty 1 increase at the beginning, but the ending value increases faster, reaches the peak value at the third loading, and then decreases rapidly. However, the initial value began to increase slightly, reached the inflection point at the second time, and then gradually decreased, but the change trend was slower than the end value. This trend reflects that the cracks in the slope gradually become larger and have a trend of overall expansion. Until the third loading, it reaches the peak value, and the slope is in a critical state, and the slope is about to lose stability and slip failure.

Feature 6 as shown in Fig.19, is the length of the longest Betty 1 barcode, and the longest Betty 1 represents the duration of the largest hole. The largest hole is the largest main crack, so it determines the final stability of the slope. It can be seen from the figure that the longest Betty 1 increases gradually from 0.4 at the initial loading, and reaches the peak at the third loading, at which time the slope is in a critical state. After the third loading, the length exceeded the maximum value of 0.75 and began to decrease gradually. This shows that the main crack reaches the maximum at the third loading, and then the slope will lose stability. This is consistent with the previous analysis of bar code.

The topological features represented by feature 8 is shown in Fig.20.

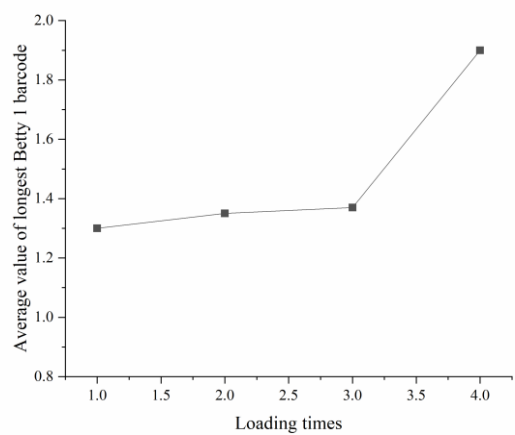

             Fig.20 Line program of feature 8

The topological feature reflected by feature 8 is the average crack width of the slope. As can be seen from the figure, the average length of the longest Betty 1 gradually increases from the initial loading, but the speed is relatively gentle. After the third loading, the curve began to rise rapidly, indicating that the slope state under the third loading was an inflection point in the evolution process. This shows that the overall crack growth of the slope is slow at the beginning, until the third loading, the slope reaches the critical state. Beyond this state, the cracks on the whole slope develop rapidly, and then instability occurs.

## 5. Conclusion

From the above simulation, it can be seen that the persistent homology feature selection based on support vector machine theory can well explain the change process and topological characteristics of slope

A new mathematical method, persistent homology, is introduced into the field of slope safety design and disaster prediction. The feasibility and accuracy of applying persistent homology to slope engineering are proved by studying the failure characteristics of slope under external load.

Based on the discrete element method, the dynamic response and crack evolution process of the slope under load are obtained. The failure characteristics of the slope are studied by using the persistent homology mathematical method. The results show that: with the persistent action of the load, the 0-dimensional Betty number changes a little, that is, the external force changes the distribution of soil particle structure, and the 0-dimensional overall presents a larger trend, which reflects that the slope gradually loses before the complete instability, and changes from 1 . It can be seen that the length of the longest Betty 1 bar code represents the size of the largest hole in the topological features, which determines the evolution characteristics and instability process of the slope, so it is the most intuitive reflection of the slope failure degree among all the characteristics.

The mutation of Betty 1 bar code means that the slope is unstable, so the slope instability can be predicted by the characteristics of the bar code. The internal geometry of the slope can be scanned by radar, which can be used to analyze the topological characteristics of the slope by the method of persistent homology, and the data can also be used to support vector machine and BP In the field of machine learning such as neural network, the intelligent and accurate research on the design and evaluation of slope protection engineering can be carried out, which provides a new idea for the research of slope safety design and disaster prediction.

## Acknowledgement

This work was supported by the Project Funded by Jiangxi Provincial Department of Science and Technology (No. 20192BBEL50028).

Journal of Computational Mechanics,2019,36(03):375-382.

[12] Pan Min, Ling Chen, Fan Jingjing. A sparse grid stochastic collocation method for slope reliability analysis[J]. Chinese Journal of Applied Mechanics, 2018,35(06):1267-1272+1419-1420.

[13] Yang T H，Wang P T，Yu Q L，et al． Effects of lateral opening angle on ore-drawing in pillarless sublevel caving based on PFC2D［C］∥International Symposium on Geomechanics and Geotechnics: From Micro to Macro.Shanghai,2010: 649-654．

[14] Zhang Jing-liang,Ju Xian-men.Application of persistent homology in image classification and recognition[J].Journal of applied mathematics and computational mathematics,2017,(4)：494-508. (in Chinese)

[15]Cang Z, Mu L, Wu K D, et al. A topological approach for protein classification[J]. Computational and Mathematical Biophysics, 2015, 3(1).